# CWBound: boundary node detection algorithm for complex non-convex mobile ad hoc networks

**Se-Hang Cheong · Yain-Whar Si**

**Abstract** Efficient message forwarding in mobile ad hoc network in disaster scenarios is challenging because location information on the boundary and interior nodes is often unavailable. Information related to boundary nodes can be used to design efficient routing protocols as well as to prolong the battery power of devices along the boundary of an ad hoc network. In this article, we developed an algorithm, CWBound, which discovers boundary nodes in a complex non-convex mobile ad hoc (CNCAH) networks. Experiments show that the CWBound algorithm is at least 3 times faster than other state-of-the-art algorithms, and up to 400 times faster than classical force-directed algorithms. The experiments also con- firmed that the CWBound algorithm achieved the highest accuracy (above 97% for 3 out of the 4 types of CNCAH networks) and sensitivity (90%) among the algorithms evaluated.

**Keywords** Mobile ad hoc network · Force-directed algorithm · Boundary node detection · Complex non-convex ad hoc network

## 1 Introduction

Applications of mobile ad hoc networks for disaster scenarios have been reported in recent years. These applications include earthquake and tsunami monitoring [1], tracking pedestrians and emergency message forwarding [2, 3, 4]. Due to their decentralized and self-organizing nature, mobile ad hoc networks are useful for instantly forming a temporary emergency network when cellular networks are over- loaded or damaged. For example, when a person is trapped under debris after an

Se-Hang Cheong
Department of Computer and Information Science, University of Macau, Avenida Da Universidade, Taipa, Macau, China.
E-mail: dit.dhc@lostcity-studio.com

Yain-Whar Si
Corresponding author, Department of Computer and Information Science, University of Macau, Avenida Da Universidade, Taipa, Macau, China.
E-mail: fstasp@umac.mo

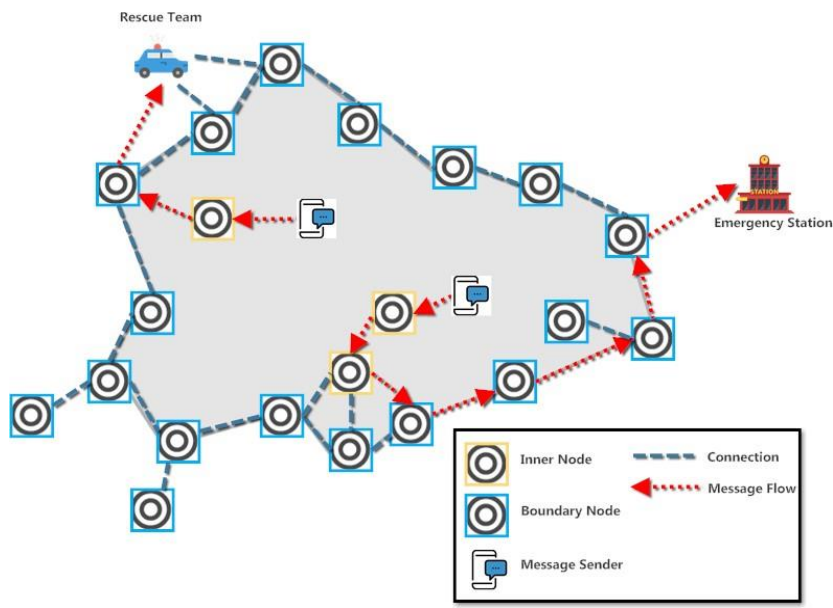

Fig. 1: A scenario for forwarding emergency messages.

earthquake, and has no cellular signal to make a call, he/she can send the messages to an emergency station via a mobile ad hoc network as illustrated in Figure 1. Search and rescue operation can be launched after the approximate location of the victim is known. Although Global Positioning System (GPS) receivers may be embedded in modern mobile phones, they are ineffective in indoor environments. Therefore, mobile ad hoc networks can be considered as a more practical solution for disaster scenarios. Mobile ad hoc networks can be formed with the Wi-Fi signals from mobile phones of the victims or battery powered wireless routers mounted in parks or residential areas where disasters are likely to occur [2]. However, mobile phones have only limited battery life and the consumption of battery power can be high when they are designed to forward the messages to nearby devices.

In the aftermath of disasters, rescuers may not be able to immediately reach the centre of the collapsed buildings or the affected areas. In these situations, boundary nodes of an ad hoc network are the kind of nodes which are more likely to establish a connection to an external emergency station or to a Wi-Fi enabled device maintained by a rescuer who is in the vicinity of the affected area. Therefore, it is essential to prolong the battery power of boundary nodes by using efficient routing protocols which maximize their throughput to store or forward messages from the inner nodes. However, location information about the boundary and interior nodes of mobile ad hoc networks is often unavailable in mobile ad hoc networks. If the boundary nodes can be identified from these networks, we can formulate strategies to prolong the battery life of these nodes by using efficient message routing protocols [5]. Moreover, with the information of boundary nodes, the location of non-boundary nodes can be estimated by using the topology information [6]. De-spite their potential use in disaster scenarios, boundary node detection problem in mobile ad hoc networks still poses a significant challenge for researchers.

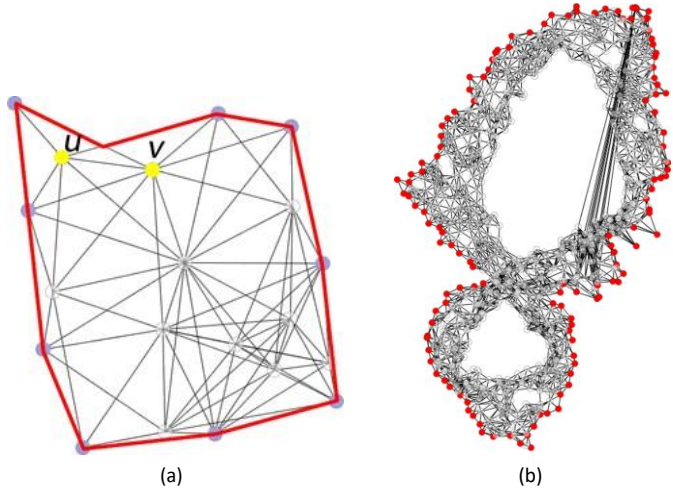

Fig. 2: (a) Illustration of boundary nodes, (b) illustration of nodes stacked on the border of a canvas.

The boundary nodes of a mobile ad hoc network are the outer nodes of a graph (network) [7]. Therefore, the boundary detection problem of a mobile ad hoc network can be reformulated as a problem of visualising the topology of the respective graph. However, the definition of boundary nodes in mobile ad hoc networks varies depending on the application of the network [8, 9, 10, 11]. In this study, we used the Mandatory Boundary Nodes [9] definition of boundary nodes. These boundary nodes are found exactly on the border of the network. Conversely, Optional Boundary Nodes are defined as nodes that are not located exactly on the border, but are positioned at a certain distance from a point or located within a defined communication range [9]. In our study, we did not consider Optional Boundary Nodes and boundary nodes of inner regions as boundary nodes. That is, we attempt to find boundary nodes exactly located on the boundary so that they have higher chance to connect to external devices deployed by rescue teams. Examples of different kinds of boundary nodes are illustrated in Figure 2(a). The red outline indicates the boundary of the network topology. Nodes $u$ and $v$ (yellow) are Optional Boundary Nodes, and purple nodes are the Mandatory Boundary Nodes considered herein. We did not consider Optional Boundary Nodes (i.e. nodes $u$ and $v$) as boundary nodes in this study.

Boundary node detection problem can be formulated as a problem of topology visualising [7]. In this paper, we use force-directed algorithms because they are widely used for the visualisation and can be used to generate a visualisation based purely on the topology information. For example, classical force-directed al- gorithms such as Fruchterman Reingold (FR) algorithm [12], the Kamada-Kawai (KK) algorithm [13] and the Davidson Harel (DH) algorithm [14] have been used for many visualisation problems. Force-directed algorithms reply on the spring force exerted on nodes and edges, which is proportional to the length of edges in the network topology. Although force-directed algorithms can be used in visual- ising the topology, several problems exist with these algorithms which, in some situations, prevent them from producing acceptable results.

In this article, we address the boundary node detection problems of complex non-convex mobile ad hoc networks (CNCAH) in which nodes are randomly dis- tributed, and only connection information is available. A CNCAH is comprised of convex and non-convex polygons, inner holes and intersecting edges. These net- works are usually non-convex in the visualisations of their topologies [15]. CNCAH networks from previous studies [7], [16] were also used in our experiments. In this article, we propose an algorithm, CWBound, which exhibits high sensitivity in generating a visualisation of the topologies of CNCAH networks. The CWBound algorithm uses a clustering based method to improve the position assignments and adjusts the forces acting on the nodes. The visualisations generated by the CWBound algorithm are then used for boundary node detection.

The rest of this article is organized as follows. In Section 2, we review related work on boundary node detection using force-directed algorithms. In Section 3, we propose CWBound algorithm for boundary node detection. In Section 4, we com- pare the CWBound algorithm with existing force-directed algorithms in solving the boundary node detection problem. In Section 5, we conclude the paper with future work.

## 2 Related Work

Boundary node detection in mobile ad hoc network has been reported in recent studies. For example, Huang et al. [17] suggested a heuristic algorithm to check if a node is suspected to be on the boundary of mobile ad hoc network in which the communication network is a $d$-quasi unit disk graph model ($d$-QUDG). Here, $d$ is the longest Euclidean distance across which two nodes can communicate. In [18], Wang et al. used topological information for boundary node and inner hole detection. The algorithm proposed by Want et al. builds the shortest path tree by flooding the network from an arbitrary root node upon initialisation. They assumed that the shortest paths selected from the shortest path tree are more similar to straight lines if there is no hole between the nodes within the shortest path tree. Otherwise, the shortest paths are curved. In addition to the heuristic approaches, topological based approaches were also used in boundary node detection. For ex- ample, in [19], Zhang et al. used Voronoi diagram [20] (a geometric data structure) for boundary node detection. In [21] and [22], geographical location data is used to identify boundary nodes and holes.

Furthermore, several extensions to force-directed algorithms have been pro- posed to solve the localisation problems of mobile ad hoc networks. For example, FRR algorithm [7] follows the approach of the classical FR algorithm, but it in- corporates information on the lengths of edges in the generation of the graph. The length information is often obtained from the signal strength of the sensors. Afzal and Beigy [23] proposed a distributed algorithm for sensor localisation based on SVM (Support Vector Machine). In their approach, each node estimates the locations according to RSS (Received Signal Strength) data and the prediction model built in SVM. The estimated locations are then sent to the sink nodes which later update the location of nodes by broadcasting to the whole network. In addition, [24] proposed a method for tracking the position of nodes using a force-directed algorithm based on the signal strength and changes in the movement of the sen- sors. In their approach, a built-in accelerometer is used to collect information from the sensors on their movements. In [25], Park et al. proposed a method to deter- mine the position of sensors based on the measured angles and distances. They assumed that each sensor can measure the distance (e.g. by using received signal strength) and the angle (e.g. by using angle of arrival (AoA) measurement) of nearby sensors.

Although force-directed algorithms can be used for boundary node detection, they suffer from high computational cost. Force-directed algorithms may not be able to generate visualisations of network topologies of acceptable quality for large- scale networks, networks with complex topologies, and networks with high average degrees [26]. In addition, sensitivity and specificity cannot be further improved when these algorithms become trapped in local minima. In addition, as we lack information about the anchor points and the coarse topology of the network, randomised coordinates are often assigned to nodes on initialisation [27]. This ran- domised position assignment causes the nodes and edges to collapse and stack against each other during the initialisation phase. Clustering based methods are useful for position assignment in initialisation phase because adjacent nodes in a mobile ad hoc network usually have similar data [28]. Therefore, in this article, we propose an algorithm called CWBound for improving the position assignments in complex networks using a clustering based method.

To evaluate the performance of the CWBound algorithm, we compare it against the KK, DH, FR, FRR, FA2 [29], and KK-MS-DS [8] algorithms. In these experiments, we measured four performance metrics: the sensitivity, specificity, accuracy, and execution time with respect to a varying number of nodes and average degrees. Experimental results showed that the CWBound algorithm shortened the processing time significantly. Specifically, we found that the CWBound algorithm was at least 3 times faster than the current state-of-the-art force-directed algorithms (for example, KK-MS-DS), and up to 400 times faster than classical force-directed algorithms (such as the KK algorithm). In most cases, the CWBound algorithm achieved 90% sensitivity, significantly greater than that of the other algorithms tested.

## 3 CWBound algorithm for boundary node detection

Force-directed algorithms can produce visualisations based purely on the structure (topology) of a network (graph) itself, and do not require additional informa- tion about the network to generate a visualisation of its topology. The classical FR algorithm models two forces on each node (attraction and repulsion). Recall that computational speed is one of the problems of classical force-directed algorithms for visualising large-scale complex networks. Moreover, nodes and edges are often collapsed, especially on the initialisation of the algorithm. Hence classical force-directed algorithms are difficult to achieve an acceptable visualisation quality within a short time. Sensitivity and specificity cannot be further improved when there are local minima. According to the evaluation conducted in [8], the sensitivity and specificity of the classical force-directed algorithm (e.g. FR) are poor, even if the number of nodes is fewer than 1000. To alleviate these prob-lems, we developed an algorithm, CWBound, which combines the clustering based schema and improved force models for boundary node detection. The combined forces in a given iteration of the CWBound algorithm can be defined as follows:

$$F = F_a + F_r + F_g \quad (1)$$

To avoid nodes and edges being stacked at the border of the canvas (Figure 2(b)), we adopted a modified version of the ideal pairwise distance from the FR algorithm. If the size of the canvas is limited, then the network has no space for expansion. Therefore, the constant of ideal pairwise distance for both the attraction ($f_a$) and repulsion forces ($f_r$) is defined in CWBound as follows:

$$k = \overline{\frac{m \times a}{n+1}} \quad (2)$$

where $m$ is an expansion multiplier for the diffusion of the nodes, $a$ is the size of canvas and $n$ is the total number of nodes in the network topology. Next, we added a gravitational force model to compensate for the repulsion of nodes that are far from the centre of the canvas. Gravitational forces attract nodes towards the centre of the canvas. This attraction can also improve node spreading so that nodes do not stack on each other. The revised gravitational force is defined as follows:

$$F_g(n) = k \times G \times d(n) \quad (3)$$

where $G$ is the gravitation coefficient, $d(n)$ is a distance function which represents the distance between the centre of the canvas and the node n, and k is the ideal pairwise distance, as defined in Equation 2.

One of the drawbacks of force-directed algorithms is the large number of iterations needed in visualisation [8]. That is because force-directed algorithms usually assign random positions to nodes at the first iteration. Although some algorithms, such as FRR, use estimated distances instead of random positions, nodes are still moved according to the forces defined in these algorithms. Sensitivity and specificity are often reduced because most of these algorithms use almost half of their execution time to construct a coarse visualisation of the topology. Therefore, a useful initialisation schema could significantly increase an algorithm's performance. Existing studies used graph partitioning [30, 31] and multilevel algorithms [32, 33] that create a coarse visualisation of network topologies. However, these algorithms depend on force-directed algorithms, to project the results of partitioning onto a canvas. The time complexity is high because we need to employ force-directed algorithms twice in which force-directed algorithms used to create a coarse visualisation and the final visualisation of network topologies.

Therefore, we proposed an initialisation scheme for nodes using a fast clustering algorithm with a suitable level of accuracy. The aim of our approach is to apply a force-directed algorithm to result of clustering results to compensate for node placement. We used clustering algorithm performs partitioning based on information regarding the nodes, edges and the weights of the edges. In our approach, we defined the estimated distance, $estdist(n_1, n_2)$, based on the free-space path loss (FSPL) assumption. This was used to calculate the estimated distances based on the signal strengths of the nodes. FSPL is a term used in telecommunications to denote the loss in signal strength of an electromagnetic wave in a line-of-sight path. The estimated distance is also used as the weight of the edges in the clustering algorithm. We assumed an ideal situation in our study; that there were no obstacles nearby to cause diffraction or reflection. The estimated distance (in meters) can be calculated for FSPL as follows:

$$d = 10^{\frac{27.55 - 20 \times log(f) - s}{20}} \quad (4)$$

where $f$ is the signal frequency in MHz and $s$ is the signal strength.

---

**Algorithm 1:** Pseudo code of CWBound algorithm

**Input :** network topology $G = (V, E)$ maximum number of execute time $s$
**Output:** a visual drawing of $G$

1 initialise the timer $t = 0$ ;
2 initialise an associated array of class C = Clustering Algorithm(G) ;
3 **while** $s > it$ **and not** $converged$ **do**
   // Clustering information
4     Let $v_{part}$ be the partition result in $C$ ;
5     Let $v_{centroid}$ be the list of centroid in $C$ ;
6     **if** distance of nodes in $v_{centroid}$ is proportional to $G$ **then**
7         $C \leftarrow Clustering\ Algorithm(G)$;
8         $v_{part} \leftarrow the partition result in C$;
9         $v_{centroid} \leftarrow the list of centroid in C$
10    **end**
      // Algorithm 2
11    CalForces( $V, E$ );
      // Update the elaspsed time
12    $t \leftarrow t + elapsed time$;
13 **end**

---

We defined a modified LinLog model as an extension of attraction force in CWBound algorithm. This force is designed to tighten the clusters so that nodes belonging to the same cluster do not separate. The idea is adopted from the FA2 algorithm. The LinLog model was proposed by [34]. The LinLog model emphasises the visualisation of clusters in a network, and tightens those clusters. A cluster may have a high number of internal edges with nodes in the same cluster, but may contain few edges with nodes outside the cluster [35]. The modified attraction force is defined as follows:

$$F_a(n_1, n_2) = log(1 + d(n_1, n_2) \times w) \quad (5)$$

where $w$ is the sum of the weight values if both $n_1$ and $n_2$ belong to the same class. d is the distance between $n_1$ and $n_2$; the initial value of $d$ is set to $est_{dist}(n_1, n_2)$. Furthermore, Equation 6 defines a new repulsion force that acts more strongly on nodes that do not belong to the same cluster, thus reducing node stacking. The repulsion force of CWBound algorithm is defined as follows:

$$F_r(n_1, n_2) = k \frac{max(C(n_1)) \times max(C(n_2))}{d(n_1, n_2)} \quad (6)$$

**Algorithm 2:** Pseudo code of forces calculation CalForces()

```
   // Calculate the displacement by repulsion force
 1 foreach u ∈ V do
 2    foreach v ∈ V and u ≠ v do
 3       u.x ← u.x + F_r(u, v);
 4       u.y ← u.x + F_r(u, v);
 5    end
 6 end
   // Calculate the displacement by attraction force
 7 foreach e ∈ E do
 8    Let u, v be the end nodes of e;
 9    Δ ← √((e.u_x − e.v_x)² + (e.u_y − e.v_y)²);
10    e.u_x ← e.u_x − (u_x−v_x)/Δ × F_a(u, v);
11    e.u_y ← e.u_y − (u_y−v_y)/Δ × F_a(u, v);
12    e.v_x ← e.v_x + (u_x−v_x)/Δ × F_a(u, v);
13    e.v_y ← e.v_y + (u_y−v_y)/Δ × F_a(u, v);
14 end
   // Calculate the displacement by gravity force
15 foreach v ∈ V do
16    Δ ← √(v²_displacementX + v²_displacementY);
17    v_x ← v_x − u_x/Δ × F_g;
18    v_y ← v_y − v_y/Δ × F_g;
19 end
```

The workflow of proposed CWBound algorithm is shown in Figure 3. On initialisation, the clustering algorithm is used to find clusters. The clustering algorithm input contains the nodes and the length of the edges. In our experiments, the lengths of the edges are randomly generated. In real-world applications, the length of an edge is an estimated distance between two nodes, and can be derived from the signal strengths of the nodes.

The next step in the CWBound algorithm is to find the centre node (medoid) of every cluster. However, at this point, none of the nodes in the network topology have positions, because they have not yet been projected onto the canvas. Therefore, the potential centre nodes can only be determined using the estimated distance. The centre node of a cluster is the node that has the shortest distances to the other nodes in the same cluster which is defined in Equation 7.

$$x_{centre} = argmin_{y \in \{x_1, x_2, \ldots x_n\}} \sum_{i=1} d(y, x_i) \tag{7}$$

where $x_1, x_2, \ldots, x_n$ are the nodes which belong to a cluster. $d(y, x_i)$ is the distance function.

Once the centre node of a cluster is identified, the network topology is projected onto a canvas by assigning randomised initial positions to nodes. Examples of the centre nodes, centroids and the distances between centre nodes are illustrated in Figure 4. The position of the centre node is then input into Equation 3. Next, the CWBound algorithm calculates the attraction, repulsion and gravitational forces using Equation 3, Equation 5 and Equation 6. The positions of the nodes are then updated iteratively. This procedure is repeated until the distances between the centre nodes are proportional to their estimated distances.

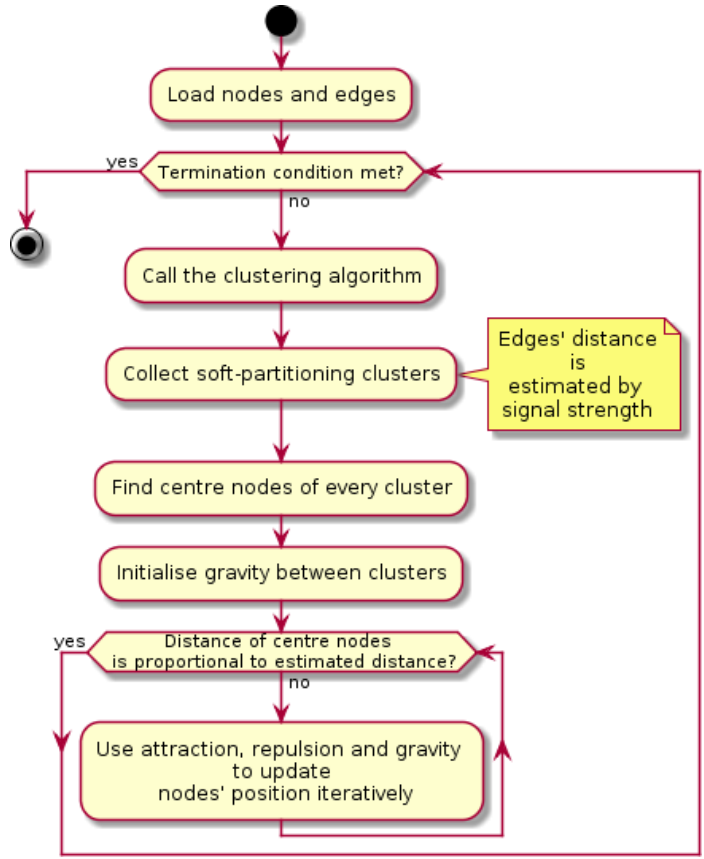

Fig. 3: The proposed CWBound algorithm.

When the distances between the centre nodes on the canvas and the estimated distances between the centre nodes are proportional, it implies that the clusters are suitably spread out and the nodes will not be stacked together, as in Figure 2(b). When this situation is detected, the following procedure is used to enhance the coarse visualisation of the topology. First, the length of the edges on the canvas are input into the clustering algorithm, instead of the estimated distances calcu- lated from the signal strength. Second, the CWBound algorithm uses Equation 8 and the centroids of the clusters to calculate the gravitational forces, rather than Equation 3. Centroids are not nodes; rather they are the positions of the centres of the clusters. The position of the centroid of a cluster is used in Equation 8. Figure 4 shows examples of centroids. We can define the gravitational forces used in this procedure as follows:

$$F_g^J(n) = k \times G \times d^J(n) \qquad (8)$$

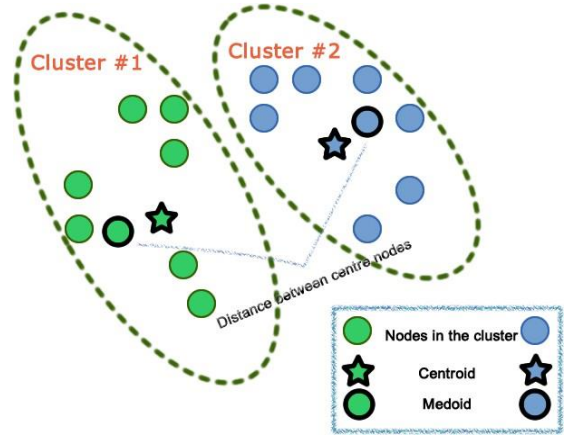

Fig. 4: Examples of centre nodes, centroids and the distances between centre nodes.

where $G$ is the gravitation coefficient, $d^{J}(n)$ is a distance function used to calculate the distance between the centroid of a cluster and a node, $n$, within that cluster, and $k$ is the ideal pairwise distance stated in Equation 2. Finally, the CWBound algorithm repeatedly executes the clustering algorithm to recalculate the centroids of the clusters and update the positions of the nodes on the canvas, until the termination condition is met. The pseudo code for the CWBound algorithm, forces calculation function and the weight calculation function are given in Algorithm 1, Algorithm 2 and Algorithm 3 respectively.

---

**Algorithm 3:** Pseudo code of weight calculation function CalWeight().

**Input :** network topology $G = (V, E)$
**Output:** Weights $W$

1 initialise the associate array of class weight $W = \emptyset$ ;
2 **foreach** $v \in V$ **and** $hopcount(u, v) = 1$ **do**
3 $\quad W[C[v]] \leftarrow 0$;
4 **end**

5 initialise the sum of class weight $sumw = 0$ ;
6 **foreach** $v \in V$ **and** $hopcount(u, v) = 1$ **do**
7 $\quad W[C[v]] \leftarrow W[C[v]] + estdist(u, v)$;
8 $\quad sumw \leftarrow W[C[v]] + sumw$;
9 **end**

  // Find a class which has maximum sum of weight.
10 Let $W_{max}$ be the maximum value of weight in $W$ ;
11 Let $C_{max}$ be the class which has highest value in $W$ ;
12 **if** $W_{max}/sumw > \delta$ **then**
13 $\quad C[u] \leftarrow C_{max}$;
14 **end**
15 **return** $W$ ;

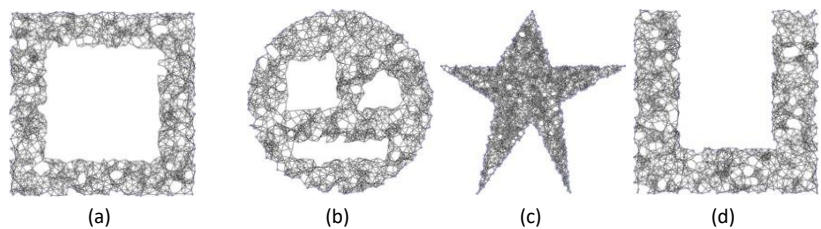

(a)　　　　　　(b)　　　　　　(c)　　　　　　(d)

Fig. 5: Samples of CNCAH networks generated for the experiments.

## 4 Experiments

In this section, we compare the performance of the proposed CWBound algorithm on CNCAH networks with the KK, FR, DH, FRR, KK-MS-DS and FA2 algorithms. These experiments were performed on a computer with an Intel Core i5 processor, 8 GB of memory, and Windows 7 64 bit. The node counts and aver- age degrees of these networks were set to 1000 and 8, respectively. Four irregular CNCAH networks were used in these benchmark topologies. Figure 5 illustrates the samples of the CNCAH networks generated for the experiments. The dataset used in our experiments can be downloaded from [1]. Besides, we have also de- veloped the CNCAHNetGenerator[2] for generating CNCAH networks of arbitrary node and edge types. In the experiments, we measured four performance metrics: the sensitivity, specificity, accuracy and execution time with respect to varying numbers of sensors and different average degrees. Definitions of these metrics are given in Table 1.

Table 1: Performance evaluation metrics.

| Metrics | Description | Result |
|---|---|---|
| Sensitivity/True positive rate | Boundary nodes on the initial network topology correctly identified as boundary nodes by algorithms | The higher the better |
| Specificity/True negative rate | Non-boundary nodes on the initial network topology correctly identified as non-boundary nodes by algorithms | The higher the better |
| Accuracy | The sum of the true positive count (i.e., boundary nodes correctly identified as boundary nodes) and the true negative count (i.e., non-boundary correctly identified as non-boundary nodes) divided by the total number of nodes examined | The higher the better |
| Execution time | Total amount of execution time that the algorithm ran in seconds | The lower the better |

---

[1] http://www.cis.umac.mo/~fstasp/tools/CWBound-Dataset.7z

[2] http://www.cis.umac.mo/~fstasp/profile eric.html

### 4.1 Evaluation of sensitivity, specificity, and accuracy in boundary node detection

We compared the sensitivity, specificity and accuracy of the CWBound algorithm in CNCAH networks with the KK, FR, DH, KK-MS-DS, FRR and FA algorithms. In this experiment, all of the algorithms were executed for 60 time units. As the results illustrated in Figure 6 show, the CWBound algorithm achieved results of at least 90% on all three measures (sensitivity, specificity and accuracy), for all of the CNCAH networks tested in the experiment. The KK-MS-DS algorithm achieved around 70% sensitivity for all of the CNCAH networks, and the remain- ing algorithms had lower sensitivities than both the CWBound and KK-MS-DS algorithms.

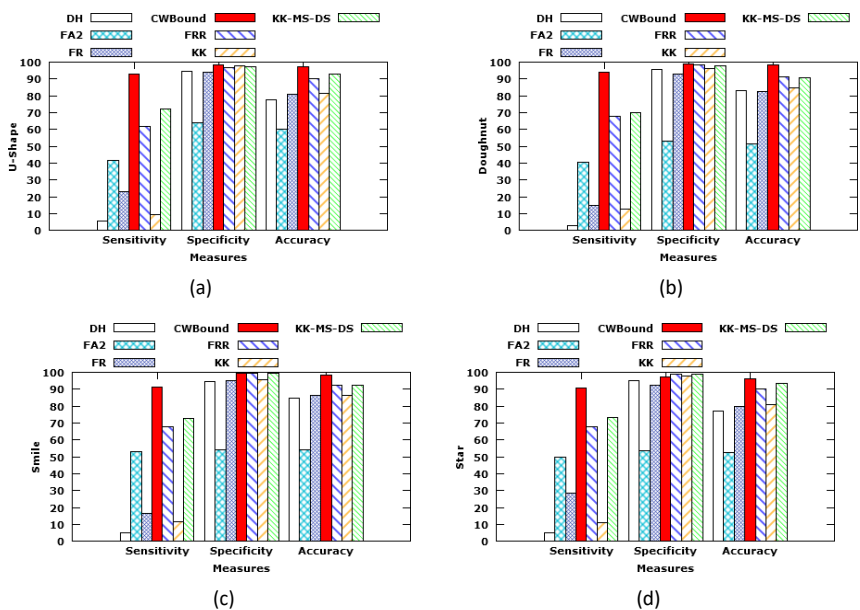

Fig. 6: Algorithm performances with (a) U-Shape, (b) Doughnut, (c) Smile and
(d) Star networks.

### 4.2 Evaluation of execution time for detecting boundary nodes

In this section, we compare the total time spent for the algorithms to achieve 90% sensitivity in the visualisations of the CNCAH networks. Two stopping criteria were set in this experiment. Either an algorithm will stop when it achieves 90% sensitivity, or the sensitivity of the algorithm remains unchanged up to certain iterations. We used 100 iterations in our experiments. Figure 7, 8, 9 and 10 sum- marise the results of our experiments. The horizontal axis of figures denotes the execution time for each network. From the experiments, we found that most of the algorithms did not achieve 90% sensitivity. The average sensitivity of the FR and FA2 algorithms was between 40% and 50% for all of the CNCAH networks evaluated.

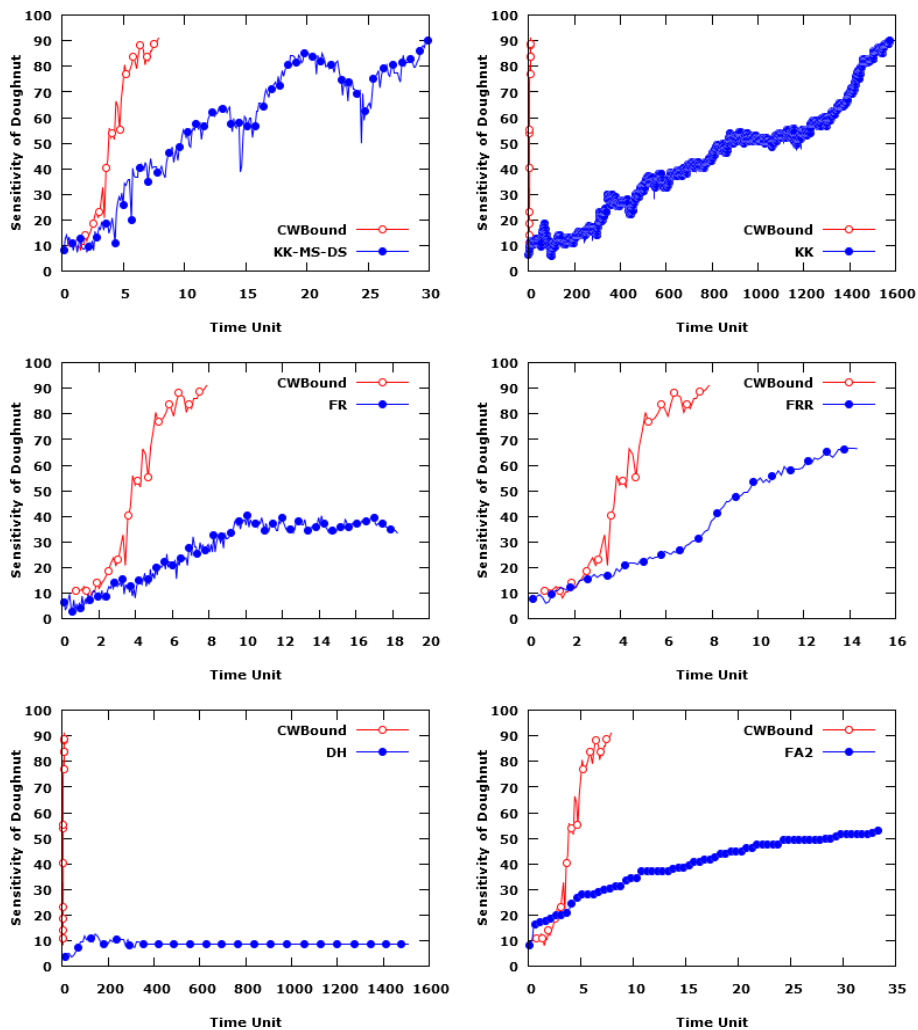

Fig. 7: Evaluation of the execution time for Doughnut CNCAH network.

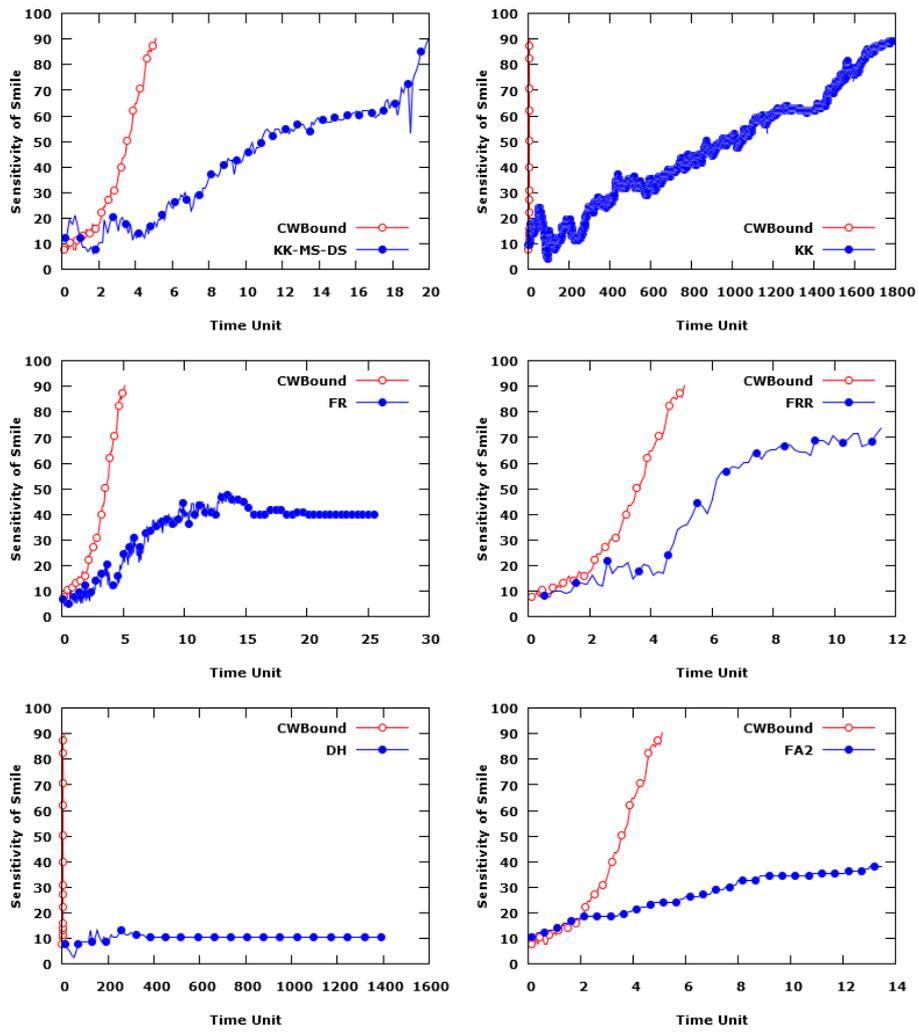

Fig. 8: Evaluation of the execution time for Smile CNCAH network.

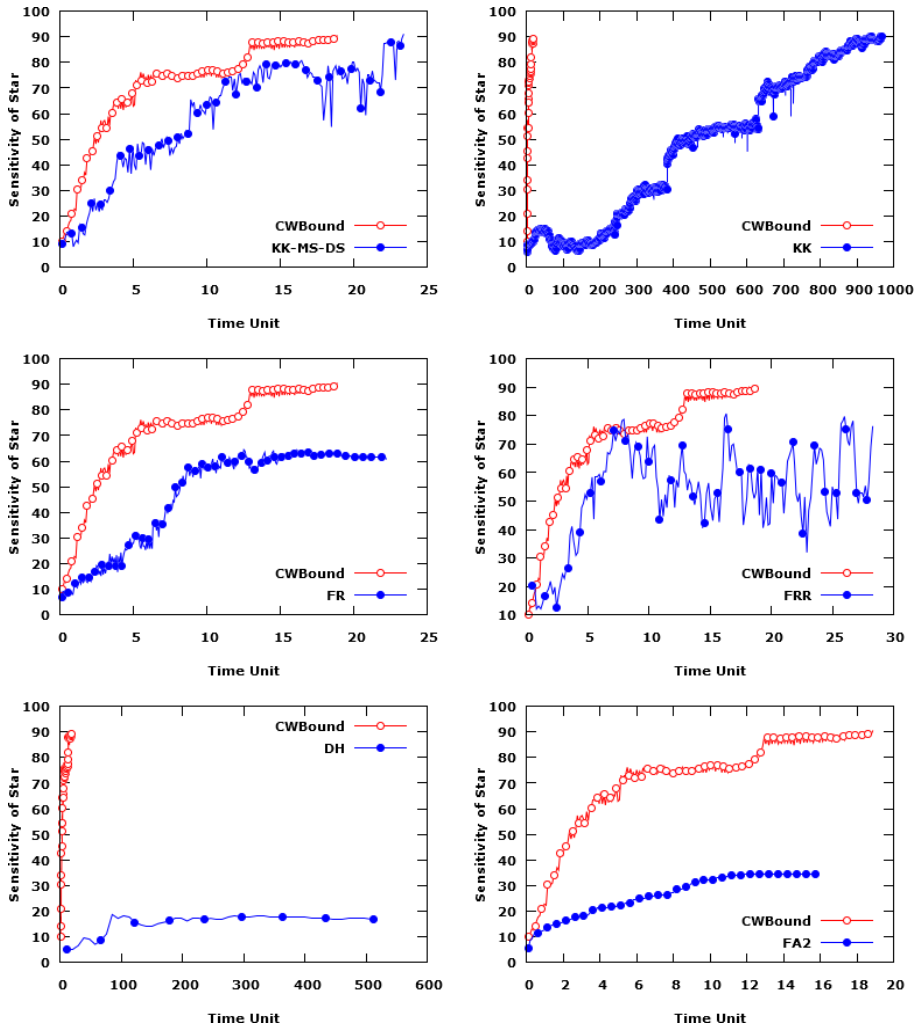

Fig. 9: Evaluation of the execution time for Star CNCAH network.

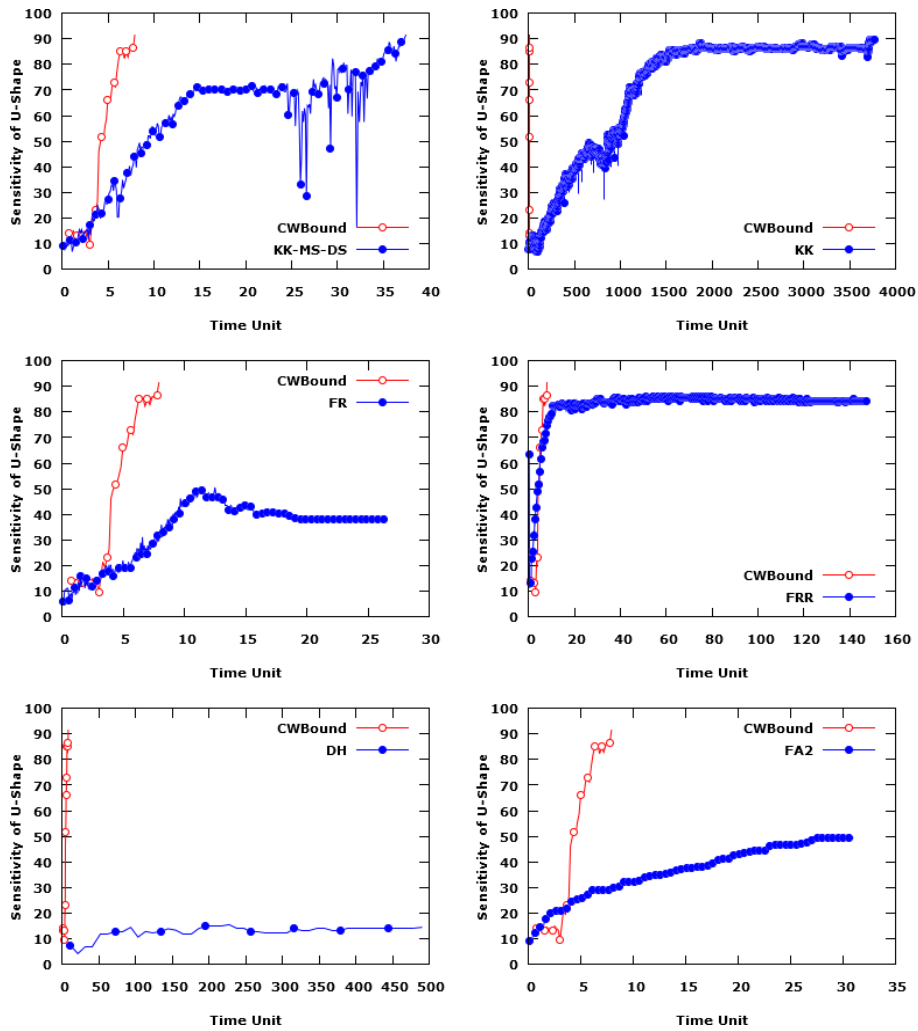

Fig. 10: Evaluation of the execution time for U-shape CNCAH network.

We also found that the DH algorithm did not reach 90% sensitivity, but remained constant at around 10%. The FRR algorithm achieved 70 to 80% sensitivity for all of the CNCAH networks evaluated. Both the KK and KK-MS-DS algorithms achieved 90% sensitivity; however, the KK algorithm exhibited the longest execution time to reach that sensitivity. From the experimental results, it is evident that the proposed CWBound algorithm reached 90% sensitivity in the shortest time.

## 5 Conclusion

The use of mobile ad hoc network in disaster scenarios has been extensively studied in recent years. By locating boundary nodes, we can design energy-efficient messaging schedules in which interior nodes can be turned off occasionally to increase the throughput. Therefore, boundary node detection is one of the crucial steps in prolonging the battery life of devices in mobile ad hoc networks for emergency situations. In this article, we proposed a novel boundary node detection algorithm called CWBound for CNCAH networks. The CWBound algorithm was evaluated against the KK, DH, FR, FRR, FA2 and KK-MS-DS algorithms. Our experimen- tal results show that the CWBound algorithm achieved the highest sensitivity of these algorithms in all benchmark CNCAH networks. The CWBound algorithm also required significantly shorter processing times. The CWBound algorithm is 3 to 5 times faster than the KK-MS-DS algorithm, and 60 to 400 times faster than the original KK algorithm in achieving 90% sensitivity in the four type of CNCAH networks (u-shape, smile, star, and doughnut) we evaluated. Our experiments also revealed that the remaining algorithms were unable to achieve 90% sensitivity. In addition, the CWBound algorithm achieved at least 95% specificity for all of the CNCAH networks evaluated, higher than any other algorithm. Con- versely, the FA2 algorithm achieved 60% specificity for u-shape, and around 50% specificity for the smile, star and doughnut CNCAH networks, which were the lowest specificities of the algorithms tested. The experiments also demonstrated that the CWBound algorithm achieved the highest accuracy (above 97% for 3 of the 4 CNCAH networks) of the algorithms evaluated.

In our future work, we plan to develop a distributed version of the proposed CWBound algorithm. The distributed version will be based on graph partitioning algorithms that separate input networks into multiple zones using soft partitioning [33, 31], and multi-level algorithms [32, 36]. In this future version, force-directed algorithms will simultaneously process every zone to produce independent visualisations of the graph topologies of each zone. These visualisations can then be aggregated into a final visualisation of a graph's topology, using the correlation information from the soft partitioning.

**Acknowledgements** This research was funded by the Research Committee of University of Macau, grants MYRG2016-00148-FST and MYRG2017-00029-FST.

## A Appendix - Chinese Whispers algorithm

We used the Chinese Whispers algorithm for clustering which is proposed by [37]. The algorithm partitions the nodes of a graph into clusters. The Chinese Whispers algorithm is able to good clustering results on large networks within an acceptable number of iterations [38]. The time complexity of the Chinese Whispers algorithm is $O(n \times E)$, where $n$ is the number of nodes and $E$ is the number of edges in the network. This algorithm uses the following steps to partition the nodes into clusters. On initialisation, every node is labelled with a random, unique, class (cluster). For example, node-1 is labelled as class #1, node-2 is labelled as class #2 and so on. In each iteration, a randomly selected node is assigned to a class in its local neighbourhood. The class selected to be assigned is the one with the highest total edge weight of all classes in the neighbourhood of the node. Step 2 is repeated until the class assigned to the nodes remains the same for a specified number ($f$) of iterations (i.e. a local minimum is reached). However, the algorithm may not always be able to converge to a local minimum. In that case, a stopping criterion is used to terminate the iterations. The pseudo code for the Chinese Whispers algorithm is given in Algorithm 4.

---

**Algorithm 4:** Pseudo code of Chinese Whispers algorithm

**Input :** network topology $G = (V, E)$ a voting
  threshold $\delta$
  maximum number of iterations $s$

**Output:** Clusters in $G$

1 initialise the iteration count $it$=0 ;
2 initialise an associated array of class $C = \emptyset$ ;
  // Step 1 - initialization of labeling.
3 initialise the id of class $x = 1$ ;
4 **foreach** $v \in V$ **do**
5   $\quad C[v] \leftarrow x$;
6   $\quad x \leftarrow x + 1$;
7 **end**

8 **while** $s > it$ **and not** $converged$ **do**
    // Step 2 - select nodes in a random order.
9   $\quad$ Let $v_{list}$ be the list of nodes $v \in V$ ;
10  $\quad v_{list} \leftarrow list\_shuffle(v_{list})$; // randomizes the order of the elements in $v_{list}$
11  $\quad$ **foreach** $u \in V$ **do**
        // Step 2.1 - Calcualte the weight of every class, the pseudo code is shown in Algorithm 3.
12  $\quad\quad$ CalWeight();
13  $\quad$ **end**
    // Step 3 - Wait until termination condition met.
14  $\quad$ **if** $C$ does not change up to certain iteration **then**
15  $\quad\quad$ The algorithm is marked as converged and end of iteration.
16  $\quad$ **end**
17  $\quad$ **else**
18  $\quad\quad$ $it \leftarrow it + 1$;
19  $\quad$ **end**
20 **end**